\documentclass[aps,prb,reprint,superscriptaddress,superscriptreference]{revtex4-1}

\usepackage{amsmath}
\usepackage{graphicx}
\usepackage{color}
\usepackage{hyperref}
\usepackage{enumitem}

\AtBeginDocument{\nocite{achemso-control}}

\newcommand{\gj}[1]{\textcolor{black}{#1}}

\begin{document}

\title{Iterative Reconstruction of Memory Kernels}

\author{Gerhard Jung}
\email{jungge@uni-mainz.de}
\affiliation{Institut f\"ur Physik, Johannes Gutenberg-Universit\"at Mainz, 
Staudingerweg 9, 55128 Mainz, Germany}
\affiliation{Graduate School of Excellence Materials Science in Mainz, Staudingerweg 9, 55128 Mainz, Germany}
\author{Martin Hanke}
\email{hanke@mathematik.uni-mainz.de}
\affiliation{Institut f\"ur Mathematik, Johannes Gutenberg-Universit\"at Mainz, 
	Staudingerweg 9, 55128 Mainz, Germany}
\author{Friederike Schmid}
\email{friederike.schmid@uni-mainz.de}
\affiliation{Institut f\"ur Physik, Johannes Gutenberg-Universit\"at Mainz, 
Staudingerweg 9, 55128 Mainz, Germany}

\begin{abstract}

 In recent years it has become increasingly popular to construct coarse-grained
 models with non-Markovian dynamics in order to account for an incomplete
 separation of time scales. One challenge of a systematic coarse-graining
 procedure is the extraction of the dynamical properties, namely the memory
 kernel, from equilibrium all-atom simulations. In this paper we propose an iterative
 method for memory reconstruction from dynamical correlation functions.
 Compared to previously proposed non-iterative techniques, it ensures
 	by construction that the target correlation functions of the original 
 	fine-grained systems are reproduced accurately by the coarse-grained system,
 	regardless of time step and discretization effects.
 Furthermore, we also propose a new numerical integrator for
 	generalized Langevin equations that is significantly more accurate
 then the more commonly used generalization of the Velocity-Verlet
 integrator. 
 
 We demonstrate the performance of the above described methods using the example
of backflow induced memory in the Brownian diffusion of a single colloid.
 For this system we are able to reconstruct realistic coarse-grained dynamics
 with time steps about 200 times larger than used in the original molecular
 dynamics simulations.

\end{abstract}

\maketitle


\section{Introduction}

The idea of systematic coarse-graining in general is going from a fine-grained model with many degrees of freedom to an effective model that is only described by a few coarse-grained variables. These coarse-grained variables are usually clusters of atomic particles. Popular examples are star polymers \cite{Hijon2010,Li2015, Li2017} or colloids \cite{Shin2010} each described by one effective particle. To understand the dynamics of coarse-grained models it is necessary to derive equations of motion for these effective degrees of freedom. In equilibrium this can be done systematically by applying the Mori-Zwanzig formalism \cite{Zwanzig1961, Mori1965, Zwanzig2001, Kinjo2007}. The formalism uses projection operators to map on the space of the coarse-grained dynamical variables. Independent of the choice of these projection operators, the resulting equations of motion can always be condensed into the form of a generalized Langevin equation (GLE), 
\begin{equation}
\mathbf{F}(t) = M \mathbf{\dot{V}}(t) = \mathbf{F}_\text{C}(t) - \int_{-\infty}^{t} \text{d}s \mathbf{K}(t-s) \mathbf{V}(s) + \partial \mathbf{F}(t)	,
\label{eq:GLE}
\end{equation}
with the coarse-grained velocities $ \mathbf{V}(t) $, mean conservative force $ \mathbf{F}_\text{C}(t) $ and random force $ \partial\mathbf{F}(t) $, respectively. The memory kernel matrix $ \mathbf{K}(s) $ contains the information of the frequency dependent responses of the coarse-grained particles. There is a deep connection between the random force autocorrelation function (RACF) and the memory kernel: the fluctuation-dissipation theorem,
\begin{equation}
\left\langle \partial\mathbf{F}(t) \partial\mathbf{F}(t') \right\rangle = k_\text{B} T \mathbf{K}(t-t') ,
\label{eq:FDT}
\end{equation}
with Boltzmann constant $ k_\text{B} $ and thermodynamic temperature $ T $. This relation can be derived using the Mori-Zwanzig formalism \cite{Zwanzig2001} or by analysis of the GLE \cite{Kubo1966} with the requirement that at equilibrium, the coarse-grained variables should be Boltzmann-distributed with respect to the conservative forces.

Derived in the 1960s  as a theoretical framework to understand the process of systematic coarse-graining, the Mori-Zwanzig formalism has become increasingly popular in recent years as a practical computational tool. The range of applications is huge: from quantum-bath oscillators \cite{Ceriotti2010} over molecular simulations \cite{Hijon2010,Carof2014a,Li2015,Li2015} to mesoscopic modeling \cite{Cordoba2012}. In the following, we will study the Brownian motion of a single colloid, therefore, the memory kernel can be considered to be a scalar function $ K(t) $ and the mean conservative force is zero, $ \mathbf{F}_\text{C}(t)=0 $.

The challenge of performing non-Markovian modeling is the extraction of the memory kernel from theory or fine-grained simulations. While the former might be generally preferable there are only few cases for which theoretical predictions are possible \cite{Guenza1999,Cordoba2012,Cordoba2012a}. Therefore, it is very important to develop efficient and precise methods to reconstruct memory kernels.

Over the years several direct reconstruction procedures have been proposed. In
principle, one can divide the existing methods into three groups: Time-based
inversion methods \cite{Shin2010,Carof2014,Lesnicki2016},
Fourier space inversion \cite{B.Schnurr1997} and parametrization techniques
\cite{Lei2016, Fricks2009}. However, these direct reconstruction techniques are sensitive to discretization errors, which may cause problems if the underlying trajectories from the microscopic model have not been generated or recorded with a very fine time resolution.
	
In the present paper, we propose a reconstruction technique that is based on an iterative procedure. This procedure allows us to optimize the memory kernel stepwise until the coarse-grained model recovers precisely the fine-grained correlation functions. The idea is motivated by the success of iterative methods in a similar inverse problem, the reconstruction of coarse-grained potentials from spatial correlation functions. The most prominent and increasingly popular methods in this field are Inverse Monte Carlo (IMC) \cite{Lyubartsev1995} and Iterative Boltzmann Inversion (IBI) \cite{Reith2003}. To the best knowledge of the authors, nobody has applied an iterative procedure to the problem of reconstructing memory kernels before.

The application of an iterative procedure requires the use of an efficient
and accurate numerical integrator for the GLE, because every error made in the integration leads to an error in the memory kernel. Additionally, the integrator should fulfill the fluctuation-dissipation theorem as precisely as possible to simulate a well-defined thermodynamic ensemble. Based on the derivation
of a Langevin integrator by Gr\o{}nbech-Jensen and Farago
\cite{Grønbech-Jensen2012} we propose in this paper an integration procedure for
the GLE that is much more accurate than the generalization of the popular
Br\"unger-Brooks-Karplus integrator (BBK) \cite{Brunger1984} 
to GLEs \cite{Li2017}.

Our paper is organized as follows: In Sec. \ref{sec:iterative_memory}, we introduce our novel iterative reconstruction technique and present an algorithm that can be implemented straightforwardly. We then derive a discretized version of the fluctuation-dissipation theorem in Sec. \ref{sec:num_integration} to propose a new numerical integrator for the GLE.  In Sec. \ref{sec:results}, we compare the iterative reconstruction to previously proposed methods using the example of the Brownian diffusion of a single colloid. The reference data is produced with molecular dynamics simulations (MD). In this section we will also show how to apply our method to increase the timescale of a coarse-grained simulation. We summarize and conclude in Sec. \ref{sec:conclusion}.


\begin{center}
	\begin{table*}
		\begin{tabular}{|c|c|c|c|}
			\hline & IBI \cite{Reith2003} & IMRF & IMRV \\ 
			\hline unknown quantity $X$ & pair potential
			$ V(r) $ & \multicolumn{2}{|c|}{memory kernel  $ K(t) $}  \\ 
			\hline matched observable $ Y $ & RDF $ g(r) $ & FACF $ \left\langle F(t) F(0)\right\rangle$ & VACF $ \left\langle v(t) v(0)\right\rangle$  \\ 
			\hline mapping function $ \phi(Y) $ &  $ 
			-\frac{1}{\beta}\ln(Y) $ & $ \beta Y $ & $ - \beta M^2 \frac{Y(t+\Delta t) - 2 Y(t) + Y(t- \Delta t)}{\Delta t^2} $ \\ 
			\hline initial guess $ X_0 $ &  \multicolumn{3}{|c|}{$ \phi(Y_\text{MD}) $} \\ 
			\hline $X=X_0$ valid in limit & $ \rho \rightarrow 0 $ & \multicolumn{2}{|c|}{$ M\rightarrow \infty $} \\ 
			\hline 
			basic iteration step & \multicolumn{3}{|c|}{$ X_{i+1} = X_{i} + \Delta \phi_i$ with $\Delta \phi_i = \phi(Y_\text{MD}) - \phi(Y_{i}) $} \\
			\hline 
			optimized iteration step & $X_{i+1} = X_i + \alpha \: \Delta \phi_i$ &
			\multicolumn{2}{|c|}{$ X_{i+1} = X_{i} + h_i(t) \: \Delta \phi_i$ (Eq.\ (\ref{eq:iteration}))} \\
			\hline 
		\end{tabular}  
		\caption{Comparison of the iterative Boltzmann inversion (IBI) 
			with the iterative memory reconstruction (IMRF, IMRV). 
			IBI reconstructs a pair potential
				from the radial distribution function (RDF) of the
				coarse-grained variables.
		}
		\label{tab:ibi_imr}
	\end{table*}
\end{center}

\section{Iterative Memory Reconstruction}
\label{sec:iterative_memory}

As an initial guess for the iterative reconstruction we will use the force autocorrelation function (FACF), because in the infinite mass limit the FACF and RACF are the same \cite{Lee2005,Theers2016},
\begin{equation}
  \lim\limits_{M \rightarrow \infty}\left\langle F(t)F(0)  \right\rangle=\left\langle \partial F(t)\partial F(0) \right\rangle = k_\text{B}T K(t)  .
\end{equation}
Additionally, also for finite masses, the memory kernel at time $ t = 0 $ is equal to the force fluctuations \cite{Zwanzig2001},
\begin{equation}
\left\langle F(0)^2  \right\rangle = \left\langle \partial F(0)^2  \right\rangle=  k_\text{B} T K(0) .
\end{equation}
These relations suggest the following iterative method, which uses the differences in the FACF between MD and GLE simulations as an additive correction to the memory kernel,
\begin{equation}
 K_{i+1}(t) = K_i(t) + \beta \left(\left\langle F(t)F(0)  \right\rangle_\text{MD} - \left\langle F(t)F(0)  \right\rangle_\text{GLE}\right) ,
 \label{eq:iteration_ansatz}
\end{equation}
with the inverse temperature $ \beta = 1/k_\text{B} T $. This procedure, the iterative memory reconstruction (IMR), is very similar to the iterative Boltzmann inversion (IBI), both of which can be seen as a simple fixed point iteration based on a zero-order approximation
\[ X \approx \phi(Y)  ,\]
where $X$ is the unknown coarse-grained quantity and $Y$ an appropriate observable; see table~\ref{tab:ibi_imr} for further details.

By Fourier transform of the iterative method and the GLE Eq.\ (\ref{eq:GLE}) one can formulate algorithm Eq.\ (\ref{eq:iteration_ansatz}) also in the frequency domain:
\begin{equation}
\hat{K}_{i+1}(\omega) = \hat{K}_i(\omega) + \beta \hat{C}_F(\omega) - \frac{{\rm{i}}\omega M\hat{K}_i(\omega)}{{\rm{i}}\omega M + \hat{K}_i(\omega)} ,
\end{equation}
with the one-sided Fourier transform of the memory kernel,
\begin{equation}
\hat{K}(\omega) = \int_{0}^{\infty} \text{d}t e^{{\rm{i}}\omega t} K(t) ,
\end{equation}
and similarly $ \hat{C}_F(\omega) $ the one-sided Fourier transform of the MD-FACF. 

In practice, the convergence of the iteration procedure described in Eq.\
(\ref{eq:iteration_ansatz}) is still poor.  When applying a global correction
we observed that differences between the FACFs \gj{did not disappear, but were}
only shifted to larger times $ t $. \gj{Introducing a constant prefactor $
\alpha < 1 $ to the iteration step (as is done in IBI) was not sufficient to
achieve a well-behaved convergence. }Therefore, the simple solution to this
problem is to introduce a ''correction'' time $ t_\text{cor} = n_\text{cor}
\Delta t  $ that localizes the time window in which the correction is applied
(see Eq.~(\ref{eq:iteration})). 
This consideration leads to the iteration prescription
	\begin{equation}
	K_{i+1}(t) = K_i(t) + h_i(t) \bigl(\phi(Y_\text{MD})-\phi(Y_{i})\bigr)
	\end{equation}
with
	\begin{equation}
	\label{eq:iteration}
	h_i(t) = \left\{ \begin{array}{ll}
	1 & t/t_\text{cor} \leq \frac{i}{2} \\
	1 - t/t_\text{cor} + i/2 &
	\frac{i}{2} < t/t_\text{cor} < \frac{i}{2} + 1\\
	0 &  t/t_\text{cor} \geq \frac{i}{2} + 1
	\end{array} \right. ,
	\end{equation}
mapping variable $Y$ and mapping function $ \phi(Y)$ as defined in
Tab.~\ref{tab:ibi_imr}. \gj{While the specific choice of $ h_i(t) $ for $t/t_\text{cor} \in [i/2,i/2+1]$ is
arbitrary, it definitely has to be nonnegative and continuous to prevent discontinuities in the memory
kernel.} The \gj{optimal} choice of $ t_\text{cor} $ is strongly system
dependent and should be \gj{determined} individually.  \gj{At small $
t_\text{cor} $, the algorithm will always converge, however, the necessary
number of iterations can be large.} \gj{ Therefore, we suggest to start with
large correction times and optimize until no ``shifting'' can be observed
anymore.} 

Another important aspect is the choice of the fine-grained correlation function used as input for the iteration. The integration procedure that will be
derived in Sec.~\ref{sec:num_integration} integrates the velocity with an error
in $ \mathcal{O}(\Delta t^3) $. However, the force can only be calculated by
finite differences,
\begin{equation}
F(t) = \gj{M} \frac{v(t+\Delta t)-v(t)}{\Delta t}  ,
\label{eq:fs_derivative}
\end{equation}
with an error in $ \mathcal{O}(\Delta t) $.  This leads to significant
deviations in the FACF at larger time steps even though the velocity
autocorrelation function $ \left\langle v(t)v(0) \right\rangle $ (VACF) is
reproduced very accurately (see e.g.  Fig.~\ref{fig:integration_comparison}).
Therefore,  we also propose a procedure that uses the VACF as matched observable
(see Tab.~\ref{tab:ibi_imr}, IMRV), motivated by the identity:
\begin{equation}
\left\langle F(t)F(0) \right\rangle = 
- M^2 \frac{\partial^2}{\partial t^2} 
\left\langle v(t)v(0) \right\rangle  .
\end{equation}

\begin{figure}
	\includegraphics{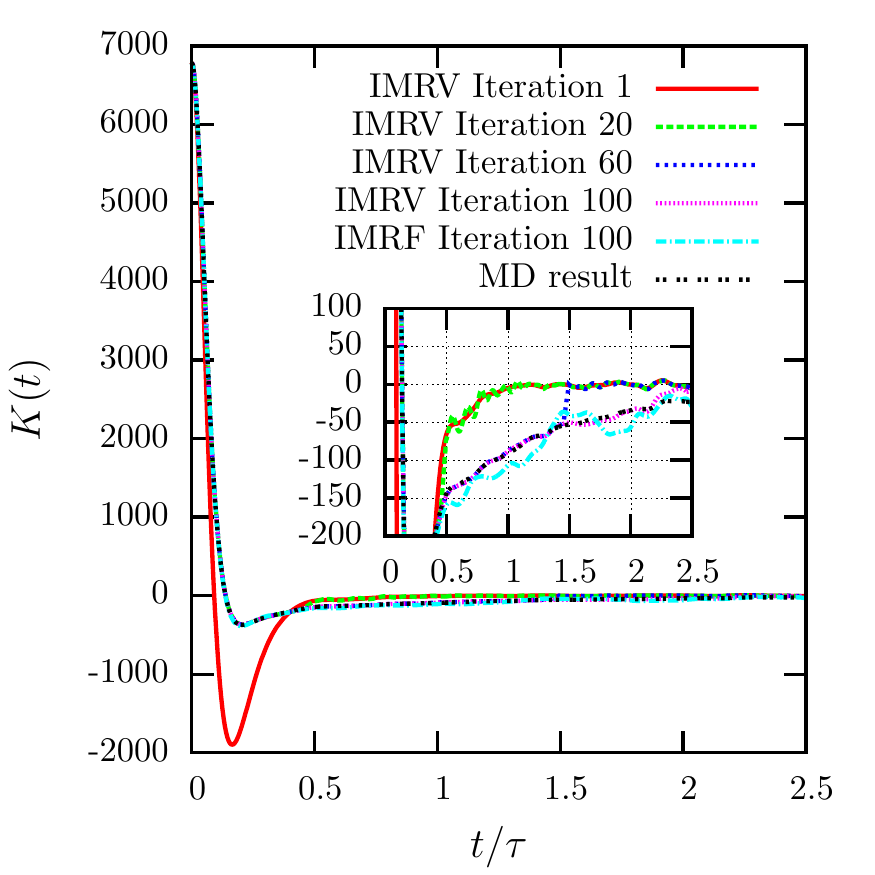}
	\includegraphics{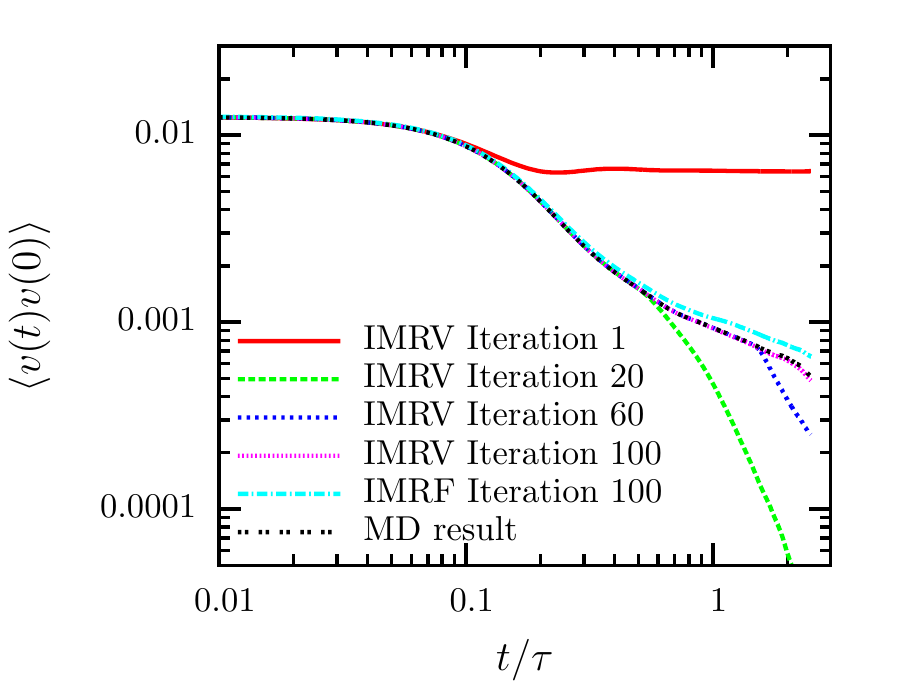}
	\caption{Iterative memory reconstruction (IMRV, \gj{IMRF}) applied to the diffusion of a single Brownian particle. In the top panel, the memory kernel is shown for several iteration steps. The ``MD result'' refers to the inverse Volterra method (see Fig.~\ref{fig:iterative_comparison}, Ref.~\cite{Shin2010}). The inset emphasizes the differences between the steps by zooming in. The bottom panel shows the VACF.}
	\label{fig:iterative_procedure}
\end{figure}

An exemplary series of iterations using the IMRV scheme is shown in Fig.~\ref{fig:iterative_procedure}. The fine-grained data were obtained from MD simulations of a single colloid in a Lennard-Jones fluid (see Sec. \ref{sec:results} for details). The figure illustrates the progress in the iterative reconstruction of the memory kernel. One can observe a well-behaved convergence that improves stepwise. The final GLE simulations \gj{using the IMRV} show no significant deviations from the MD simulations. \gj{Furthermore, one can see that for this time step, the IMRF gives accurate results as well.}

To apply this procedure to GLE simulations with larger time steps we need to find an accurate numerical integrator for the GLE in Eq.~(\ref{eq:GLE}), because every error made in the discretization will lead to a similar error in the memory kernel. Such an integrator will be proposed in the next section.

\section{Numerical Integration of the GLE}
\label{sec:num_integration}

To derive a numerical integrator for the GLE it is necessary to find discretized versions of both Eq.~(\ref{eq:GLE}) and the fluctuation-dissipation theorem Eq.~(\ref{eq:FDT}). The first step in discretizing the GLE is to introduce a discrete memory kernel,
\begin{equation}
M \dot{v}(t)= - \sum_{m=0}^{N-1} K_m v(t-m \Delta t) + \partial F(t)  ,
\label{eq:GLE_discrete}
\end{equation}
with a not yet specified memory sequence $ K_m $ with $ m = 0,...,N-1 $, and a finite time step $ \Delta t $.

To derive the fluctuation-dissipation theorem we first identify the frequency dependent response $ \hat{\gamma}(\omega) $ of the discretized memory kernel by Fourier transform of Eq.~\ref{eq:GLE_discrete},
\begin{eqnarray}
\hat{v}(\omega) &=& \frac{\partial\hat{F}(\omega)}{{\rm{i}}\omega M + \hat{\gamma}(\omega)} \quad , \\
\text{with } \hat{\gamma}(\omega) &=& \sum _{m=0}^{N-1} K_m e^{-{\rm{i}}wm\Delta t}  .
\end{eqnarray}
Here, $ \hat{v}(\omega) $ denotes the Fourier transform of the velocity,
\begin{equation}
\hat{v}(\omega) = \int_{-\infty}^{\infty} \text{d}t e^{{\rm{i}}wt} v(t)  ,
\end{equation}
and similarly $ \partial\hat{F}(\omega) $ the Fourier transform of the random force. Following the derivation of Hauge and Martin-L\"{o}f \cite{Hauge1973} for the continuous version of the GLE, we can identify the power-spectrum $ \hat{C}_{\partial F}(\omega) $ of the random force, defined as the Fourier transform of the RACF,
\begin{equation}
\hat{C}_{\partial F}(\omega) = 2 k_B T \Re \{ \hat{\gamma}(\omega) \} = 2 k_B T \sum _{m=0}^{N-1} K_m \cos(wm\Delta t)  ,
\end{equation}
with the real part $ \Re\{\hat{\gamma}(\omega)\} $ of the response function. By inverse Fourier transform we finally find the fluctuation-dissipation theorem,
\begin{equation}
\left\langle \partial F(t) \partial F(t') \right\rangle = k_B T \sum _{m=0}^{N-1} a_m K_m \delta(t-t'-m\Delta t)  ,
\label{eq:fdt_discrete}
\end{equation}
with $ a_0 = 2 $ and $ a_m = 1 \text{ for } m\neq0 $.
The most important message of this equation is the factor of 2 that enters the instantaneous contribution of the correlation function. This factor is also present in the fluctuation-dissipation theorem for the Langevin equation without memory.

On the basis of this derivation we now develop a numerical integrator for the GLE. The scheme is inspired by the derivation of a Langevin integrator by Gr\o{}nbech-Jensen and Farago \cite{Grønbech-Jensen2012}.

The first step is to integrate Eq.~(\ref{eq:GLE_discrete}) over a time interval $ \Delta t $ between the times $ t_n $ and $ t_{n+1} = t_n + \Delta t $:
\begin{equation}
\int\limits_{t_n}^{t_{n+1}} M \dot{v} \text{d}t' = - \sum _{m=0}^{N-1} K_m \int\limits_{t_n}^{t_{n+1}} \dot{r}(t'-m \Delta t) \text{d}t' + \int\limits_{t_n}^{t_{n+1}} \partial F(t') \text{d}t' \,\, .
\end{equation}
This can, without approximation, be written as,
\begin{equation}
v_{n+1}-v_{n} = - \sum _{m=0}^{N-1} \frac{K_m}{M} (r_{n+1-m}-r_{n-m}) + \frac{1}{M} \beta_{n} ,
\end{equation}
with the velocity $ v_n = v(t_n) $, position $ r_n = r(t_n) $ and correlated Gaussian distributed random numbers  
\begin{equation}
\left\langle \beta_{n+m} \beta_{n} \right\rangle = k_B T a_m K_m \Delta t  .
\end{equation}  

To find a closed form for this integrator we need to approximate the relation $ \dot{r} = v $ by the term 
\begin{equation}
r_{n+1}-r_{n} \approx \frac{\Delta t}{2}(v_{n+1}+v_n)  .
\end{equation}
This introduces an error in the algorithm that scales with $ \Delta t^3 $ and is therefore the limiting factor of this discretization scheme. Formerly derived schemes for Langevin or Brownian dynamics are based on similar approximations (e.g. Ref.~\cite{Grønbech-Jensen2012,Thalmann2007,Ricci2003}). With these relations we can finally write down the full integration algorithm for the discretized GLE in Eq.~(\ref{eq:GLE_discrete}):
\begin{eqnarray}
r_{n+1} &=& r_n + b\Delta t v_n - \frac{b \Delta t}{2} f^\text{D}_n + \frac{b \Delta t}{2 M} \beta_{n}\nonumber\\
v_{n+1} &=& a v_n - b f^\text{D}_n + \frac{b}{M}\beta_{n}  ,
\label{eq:gle_integrator}
\end{eqnarray}
with
\begin{eqnarray}
f_n^\text{D}&=&\sum _{m=1}^{N-1} \frac{K_m}{M} (r_{n+1-m}-r_{n-m})  ,\label{eq:dissipartive_force}\\
a &\equiv& \frac{1-\frac{K_0 \Delta t}{2M}}{1+\frac{K_0 \Delta t}{2M}} \qquad  \text{ and } \qquad b \equiv \frac{1}{1+\frac{K_0 \Delta t}{2M}}  .
\end{eqnarray}

In the special case of $ K_m = 0 \text{ for } m\neq0 $, these equations reduce to the GJF-integrator proposed in Ref. \cite{Grønbech-Jensen2012}. 

To test the algorithm, we use the memory kernel that was reconstructed in Sec. \ref{sec:iterative_memory}. The discretization of the memory kernel will be chosen as follows
\begin{equation}
K_m = \begin{cases}
\frac{K(0) \Delta t}{2} \text{ for } m=0\\
K(m\Delta t) \Delta t \text{ for } m \neq 0
\end{cases}
\label{eq:memory_discrete}
\end{equation}
This choice is consistent with the continuous version of the GLE because firstly it leads to a proper discretization of the integral (trapezoidal rule) and secondly it recovers a random force autocorrelation function that is similar to the continuous memory kernel due to Eq.~(\ref{eq:fdt_discrete}).

The colored noise was produced using the Fourier transform technique proposed by Barrat \emph{et al.} \cite{Barrat2011,Li2015} (see App.~\ref{ap:noise}). We will compare the results to a generalization of the popular Br\"unger-Brooks-Karplus integrator \cite{Brunger1984,Li2017}, using the discretization
\begin{eqnarray}
r_{n+1} &=& r_n + v_n \Delta t + \frac{\Delta t^2}{2M}f^n \nonumber\\
v_{n+1} &=& v_n+\frac{\Delta t}{2M}(f_n+f_{n+1}) \nonumber\\
 \qquad f_n &=& - \sum_{m=0}^{N-1} K_m v_{n-m} + \partial F_{n}  ,
 \label{eq:vv_integrator}
\end{eqnarray}
with $ \partial F_{n} = \partial F(t_n) $. 

The observables that will be determined are the reduced temperature $ k_\text{B} T = M \left\langle v(0)^2 \right\rangle  $ and the VACF for two integration time steps $ \Delta t = 0.001 $ and $ \Delta t = 0.1 $.

\begin{figure}
\includegraphics{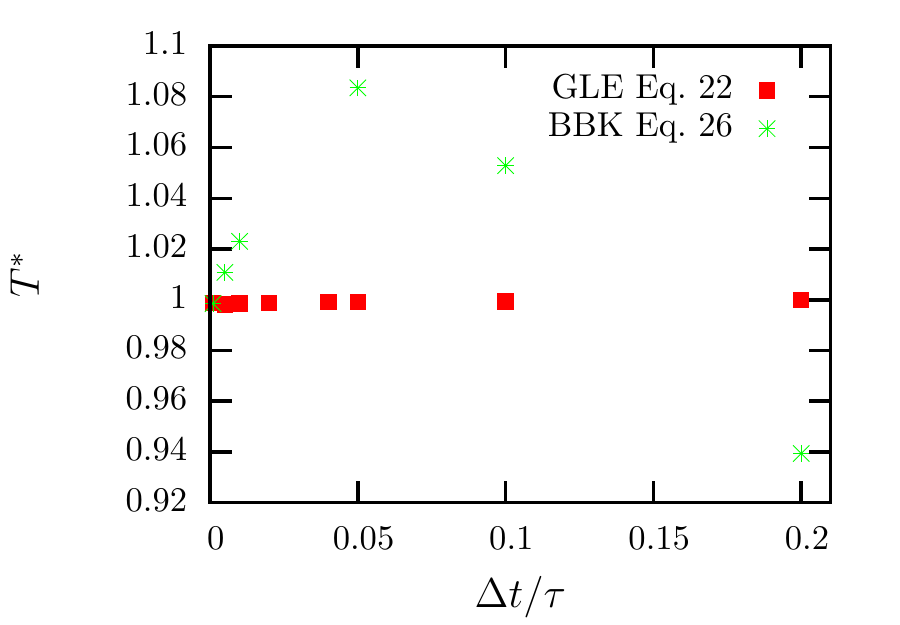}
\includegraphics{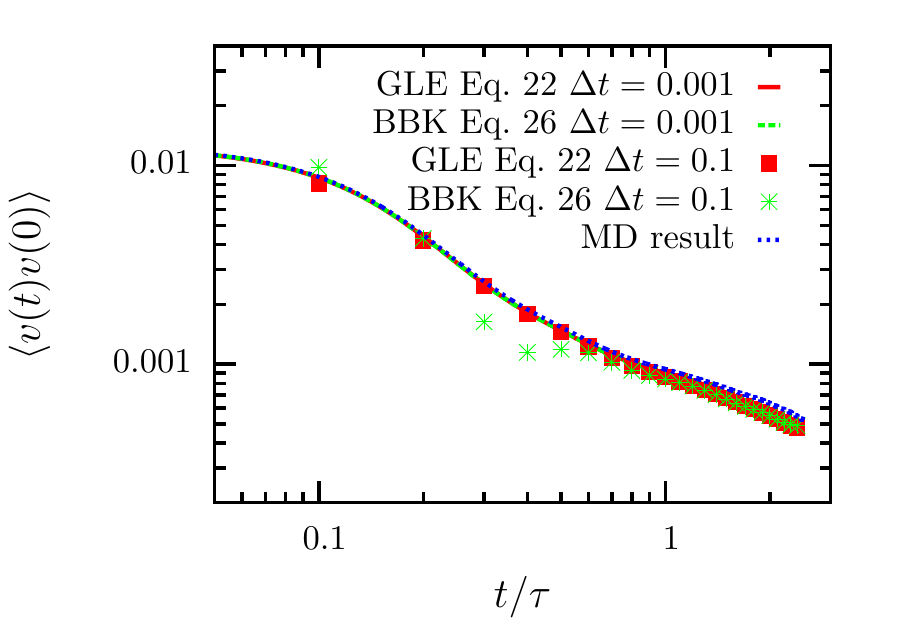}
\caption{Reduced temperature $ T^* $ (top panel) and VACF (bottom panel) of GLE simulations with different time steps. The target temperature is $  T^*_0 = 1 $. An illustration of the inserted memory kernel can be found in Fig.~\ref{fig:iterative_comparison}. }
\label{fig:integration_comparison}
\end{figure}

In Fig. \ref{fig:integration_comparison} we can observe that with the new integration scheme the kinetic energy is correct even for very large time steps. In contrast, the BBK integrator shows huge deviations from the target temperature (Fig.~\ref{fig:integration_comparison}, top panel). This observation is supported by the VACF (Fig.~\ref{fig:integration_comparison}, bottom panel). While for small time steps the results of both integrators agree with MD, there are significant deviations when using the BBK integrator at larger time steps. On the contrary, the proposed GLE integrator still shows remarkably good agreement even for time steps that are 100 times larger than the ones used in MD.


\section{Results}
\label{sec:results}

With the iterative memory reconstruction and the integrator established, we can now apply them to the diffusion of a single colloid in a Lennard-Jones (LJ) fluid. 

The system we considered was created by placing LJ particles on a fcc-lattice with lattice constant $ a = 1.71 \sigma $ and therefore a reduced density of $ \rho^* = \rho \sigma^3 = 0.8 $. The reduced temperature was set to $ T^* = k_\text{B} T / \epsilon = 1.0  $. The LJ diameter $ \sigma $, energy $ \epsilon $ and time $ \tau = \sigma \sqrt{m / \epsilon} = 1 $ are defining the length, energy and time units of the simulation. To calculate the interactions we chose a LJ cutoff $ r_\text{c} = 2.5 \sigma $ and a particle mass $ m^*= 1m $. The colloid was carved out off the fcc-lattice with a radius $ R = 3 \sigma $, and then defined as rigid body so that the inter-particle distances were fixed. The resulting colloid mass was $ M = 80 m $ with a hydrodynamic radius $ R_\text{H} = 2.7 \sigma $ (determined by the radial distribution function). The cubic simulation box had a size of $ L = 41.04 \sigma $ with periodic boundary conditions in all three dimensions. To sample at the correct temperature, we equilibrated the system using a Langevin thermostat. The system was then integrated with a time step of $ \Delta t_\text{MD} = 0.001 $ in the \emph{NVE}-ensemble. The simulations were performed with the simulation package \emph{Lammps} \cite{Plimpton1995}. We use this system as toy model to analyze the performance of different memory reconstruction techniques. Therefore, we will not systematically study finite size effects or the radius dependence of the memory kernel. Nevertheless, we verified that the data are in quantitative
	agreement with the predictions of hydrodynamic theory \cite{Theers2016} 
(curves not shown here). 

In the following, we compare different memory reconstruction techniques: (i) Backward orthogonal dynamics \cite{Carof2014,Lesnicki2016} to first and second order (see also App.~\ref{ap:2nd}), (ii) the inverse Volterra method \cite{Shin2010} both with a time discretization $ \Delta t_\text{MD} = 0.001 $ and (iii) the IMRV method with a time step $ \Delta t_\text{GLE} = 0.005 $ and a correction time $ t_\text{cor} = 0.05 $. In the numerical simulation, the memory kernel was evaluated until a cutoff time $ t_\text{cut} = 2.5 $ resulting in a memory sequence $ K_m $ of 500 elements.

\begin{figure}
\includegraphics{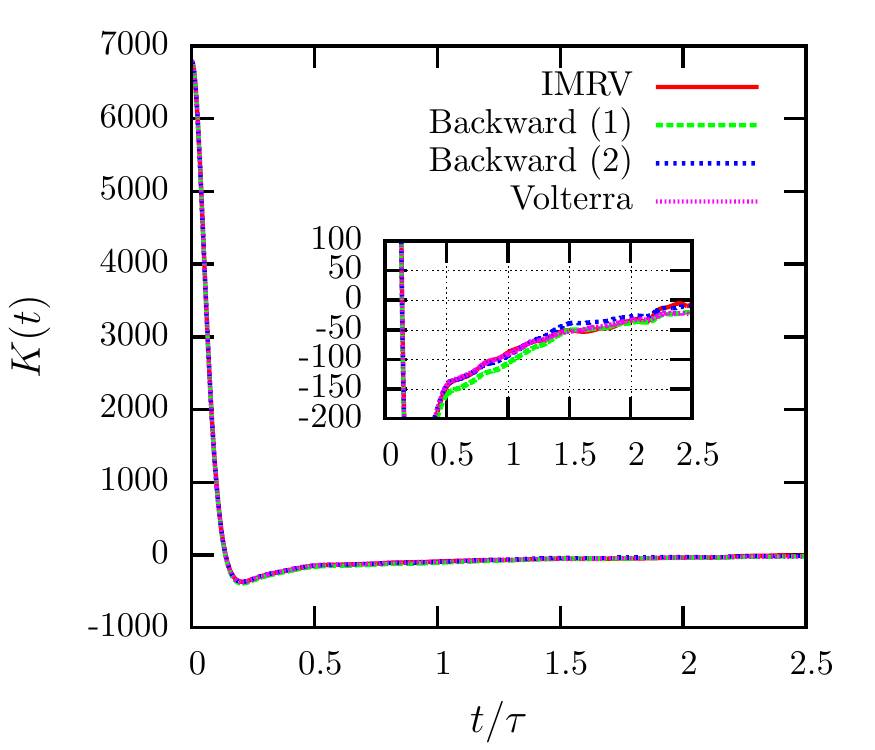}
\includegraphics{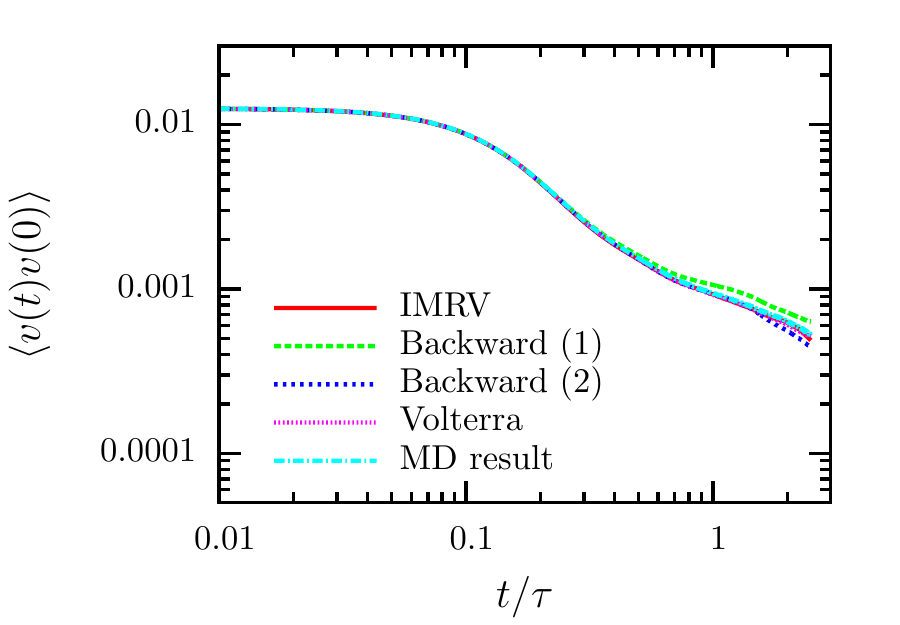}
\caption{Memory kernel (top panel) and VACF (bottom panel) of a single Brownian particle as described in Sec.~\ref{sec:results}. The kernel was reproduced using the iterative memory reconstruction (IMRV), the Backward orthogonal dynamics to first order \cite{Carof2014} and second order \cite{Lesnicki2016} and the inverse Volterra method \cite{Shin2010} with a time discretization $ \Delta t_\text{MD} = 0.001 $.}
\label{fig:iterative_comparison}
\end{figure}

The results for the memory kernel are illustrated in Fig.~\ref{fig:iterative_comparison} (top panel). The figure shows that all methods reproduce a similar memory kernel and therefore the same dynamical properties. This impression is supported when comparing the VACFs produced in GLE simulations using the different memory kernels (see Fig.~\ref{fig:iterative_comparison}, bottom panel). Only the first order Backward orthogonal dynamics shows small deviations from the MD results.

\begin{figure}
\includegraphics{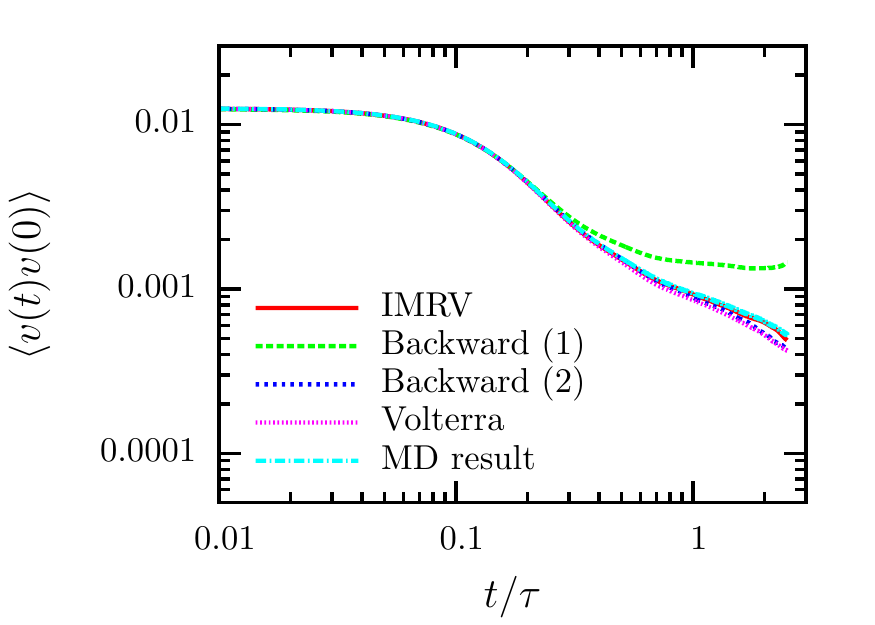}
\caption{VACF of a single Brownian particle as described in Fig.~\ref{fig:iterative_comparison}. The time discretization was $ \Delta t_\text{MD} = 0.005 $.}
\label{fig:iterative_comparison_large}
\end{figure}

To test the time discretization effects in the considered methods, we reduced the discretization of the reconstruction to $ \Delta t_\text{MD} = 0.005 $. Fig.~\ref{fig:iterative_comparison_large} confirms the observations we made in the previous paragraph. The first order Backward orthogonal dynamics suffers from a significant time step dependence that was already pointed out in Ref.~\cite{Carof2014}. \gj{However, the accuracy of the Backward orthogonal dynamics can be very much improved by applying the second order scheme.} The figure also suggests a small time step dependence of the inverse Volterra method but nonetheless the results are still fairly accurate. Compared to the other methods, the IMRV methods clearly performs best, the results are almost indistinguishable from the original MD data. 

\begin{figure}
\includegraphics{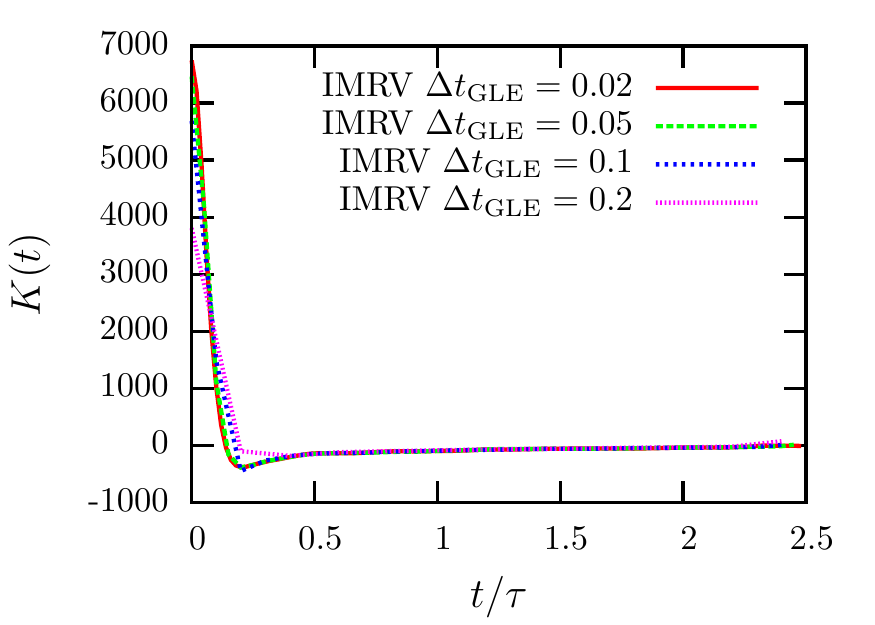}
\includegraphics{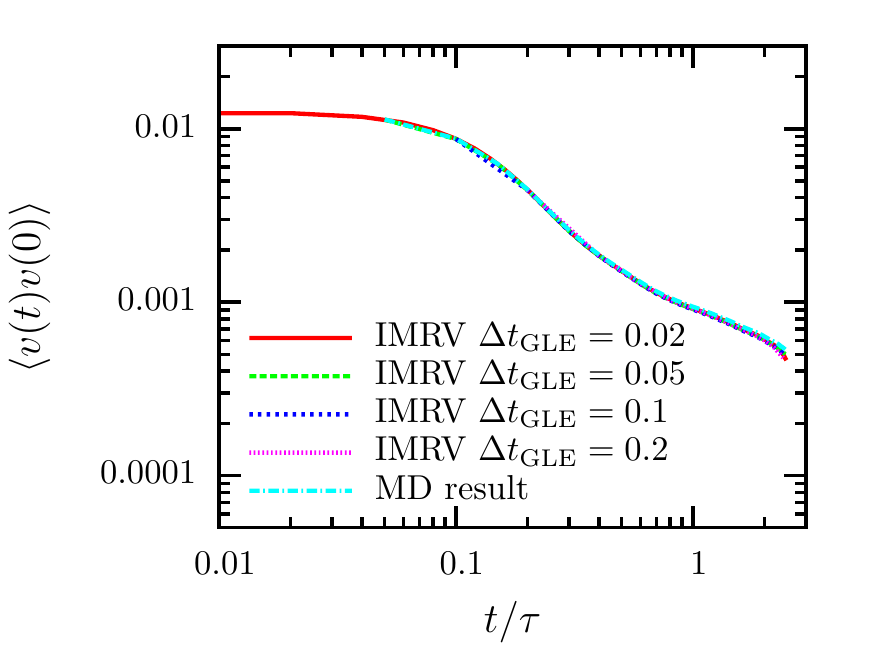}
\caption{Memory kernel (top panel) and VACF (bottom panel) of a single Brownian particle as described in Sec.~\ref{sec:results}. The iterative memory reconstruction (IMRV) was applied for various different integration time steps. \gj{The results of the IMRV are more accurate, than the straightforward integration of the GLE using the original memory kernel (see Fig.~\ref{fig:integration_comparison}).}}
\label{fig:iterative_large_timesteps}
\end{figure}

 We now want to test the limits of the IMR and apply it with discretization time steps $ \Delta t_\text{GLE} = 0.02 - 0.2 $. The correction time was chosen to be $ t_\text{cor} = 2 \Delta t_\text{GLE} $. For the largest time steps we only updated the correction window (see Eq.~\ref{eq:iteration}) every second or fourth iteration to approximately conserve an effective correction time $ t_\text{cor} = 0.05 $. Surprisingly, the \gj{IMRV} converged for all considered time steps (see Fig.~\ref{fig:iterative_large_timesteps}). The discretized memory kernels are naturally different for larger time steps because each value represents effectively the average of the continuous memory kernel over a broader time window. \gj{In the above problem the IMRF did not converge for larger time steps. The reason is the already mentioned discretization error in the force determination (see Eq.~\ref{eq:fs_derivative}). }
 
These studies show the strength of the iterative reconstruction. Independent of the time step it can always find a memory kernel that reproduces precisely the desired VACF (or FACF). This is a very noteworthy and important result because it enables us to significantly increase the time step of the coarse-grained 
simulations. In practice, this means that the choice of the time step
	will not be limited by the dissipative and stochastic part of the GLE in most
	cases, but by the conservative part.  Since the computational time of the
algorithm scales approximately linear with the time step (because the number of
elements in the sum in Eq.~\ref{eq:dissipartive_force} is reduced) the
efficiency of the sampling scales quadratically. Therefore, the proposed
coarse-graining of the timescale is essential for the applicability and
efficiency of the generalized Langevin equation.


\section{Conclusion}
\label{sec:conclusion}

In this paper we proposed a new technique to reconstruct memory kernels from
molecular dynamics simulations. Instead of using direct inversion techniques,
we developed an iterative algorithm to determine memory kernels that precisely
reproduce the fine-grained dynamics. \gj{We showed that this procedure opens up
an elegant way of finding discretizations of memory kernels that reproduce
realistic dynamics even for very large time steps. The consequence is that the
size of the time step for physical models will be restricted by the
conservative forces, and not by the resolution of the memory kernel.}
Additionally, the strength of the method is its flexibility in optimizing one
target observable (e.g. VACF or FACF), instead of being restricted to a
predetermined inversion. This could grant some freedom to increase the
representativity and transferability of the coarse-grained model, similar to
related work concerning IBI (e.g. Ref. \cite{Villa}). \gj{In particular,}
\gj{the IMR could allow the construction of a multi-criteria algorithm by
optimizing a weighted combination of several time correlation functions.}

To use the iterative reconstruction with good accuracy, we proposed a novel discretization scheme for the GLE. The tests suggest that the new integrator performs much better than a generalization of the popular BBK integrator, and that the velocity is integrated accurately even for very large time steps. Additionally, we derived a discretization of the fluctuation-dissipation theorem for the GLE. We can show that the discretized version counters intuition and introduces a scaling factor of 2 to the instantaneous fluctuations, similar to the Langevin equation. This result is often overlooked and needs to be generally considered when performing simulations with the generalized Langevin equation.

With the present work as foundation, we plan to study more complex systems of
interacting colloids. This will introduce pair potentials and pairwise memory
kernel, similar to Refs.~\cite{Li2015,Li2017}.  It
will be interesting to investigate whether we can combine iterative methods to
determine pair potentials with the iterative memory reconstruction. Additionally, we plan to systematically analyze the mathematical foundations of the iterative algorithm to gain a deeper understanding of its properties and convergence.  
\vspace*{0.4cm}

\section*{Acknowledgment}

The authors want to thank Jay D. Schieber and Oded Farago for helpful discussions. This work was funded by the German Science Foundation within project A3 of the SFB TRR 146. Computations were carried out on the Mogon Computing Cluster at ZDV Mainz.


\appendix

\section{Colored noise}
\label{ap:noise}

\gj{ The goal is \gj{to generate} correlated Gaussian distributed random numbers $ \beta_n $, \gj{that fulfill} the FDT, 
\begin{equation}
\left\langle \beta_{n+m} \beta_{n} \right\rangle = k_B T a_m K_m \Delta t  ,
\end{equation}  
 with the discretized memory sequence $ K_m $ for $ m=0,..., N-1 $ and the parameter $ a_m $ defined by $ a_0 = 2 $ and $ a_m = 1 \text{ for } m\neq0 $ (see Sec.~\ref{sec:num_integration}).}

\gj{ First, we introduce the real parameter $ \alpha_s $ for $ s=-N+1,..., N-1 $, defined by,
\begin{equation}
a_m K_m \equiv \sum_{s=-N+1}^{N-1} \alpha_s \alpha_{s+m},
\label{eq:noise_ansatz}
\end{equation}
where $ \alpha_{s+m} = \alpha_{s+m-2N+1} $ if $ s+m \geq N $. It can now be shown that for a sequence of uncorrelated Gaussian distributed random numbers $ W_n $, the relation,
\begin{equation}
\beta_n = \sqrt{k_B T \Delta t} \sum_{s=-N+1}^{N-1} \alpha_s W_{n+s}, 
\end{equation}
generates random numbers with the target correlation function. Therefore, the challenge is \gj{to determine} the parameter $ \alpha_s $. This can be achieved by applying the discrete Fourier transform (DFT) on the memory sequence,
\begin{equation}
\hat{K}_k = \sum_{m=-N+1}^{N-1} a_m K_m e^{-{\rm i}km \frac{2 \pi }{2N-1} },
\end{equation}
with $ K_{-m} = K_m $. We will now define
\begin{equation}
\hat{\alpha}_k =  \sqrt{\hat{K}_k}
\label{eq:noise_def}
\end{equation}  
and consequentially determine the parameter $ \alpha_s $ by inverse DFT
\begin{equation}
\alpha_s = \frac{1}{2N-1} \sum_{k=-N+1}^{N-1} \hat{\alpha}_k e^{{\rm i}ks \frac{2 \pi }{2N-1} }.
\end{equation}
It is straightforward to show that the definitions in Eqs.~(\ref{eq:noise_ansatz}) and (\ref{eq:noise_def}) are consistent.}

\gj{ \gj{We close with one important comment. In practical applications, a small number of coefficients $\hat{K}_k$ was slightly below zero.} While this
is theoretically impossible, because the modes connected to the negative values
$ \hat{K}_k $ would be instable, the problem can occur due to discretization
errors if the memory sequence $ K_m $ has ``edges''. The solution used in this
paper is zeroing all values $ \hat{K}_k < 0 $. Although, this leads to very
small deviations from the FDT theorem ($ \approx 0.01 \% $), the difference did
not have any physical consequence.}

\section{Backward orthogonal dynamics}
\label{ap:2nd}

\gj{In Ref.~\cite{Carof2014} Carof \emph{et al.} introduced the idea to
calculate the memory kernel by reconstruction of the random force $ \partial
\mathbf{F}(t) $ using molecular dynamics simulations. Their method is directly
based on the Mori-Zwanzig formalism.} \gj{In the context of the present work,
we have extended their Backward orthogonal dynamics scheme to second order.
The derivation is presented in the following.}

\gj{The memory kernel can be calculated using the relation,
\begin{equation}
K(t) = \left\langle \sum_{i} \tilde{F}_i^{\overline{t}} F_i(t)  \right\rangle,
\end{equation}
with the force $ F_i(t) = M_i \dot{v_i}(t) $ and the projected force $ \tilde{F}_i^{\overline{t}} $ on particle $ i $. The projected force is defined by
\begin{equation}
\tilde{F}_i^{\overline{t}} = F_i(0) + \int_{0}^{t} \text{d}u F_i(u) \frac{\left\langle M_i v_i(u) \tilde{F}_i^{\overline{u}} \right\rangle }{\left\langle M_i^2 v_i(0)^2 \right\rangle }.
\label{eq:PF}
\end{equation}
These equations correspond to Eqs.~(B2) and (B5) in Ref.~\cite{Carof2014} with $ A(t) = B(t) = F_i(t) $. To reconstruct the dynamics of the projected force $ \tilde{F}_i^{\overline{t}} $ we rewrite Eq.~\ref{eq:PF} into a discretized integration scheme:
\begin{equation}
\tilde{F}_i^{\overline{t+\Delta t}} = \tilde{F}_i^{\overline{t}} + \int_{t}^{t+\Delta t} \text{d}u F_i(u) \frac{\left\langle M_i v_i(u) \tilde{F}_i^{\overline{u}} \right\rangle }{\left\langle M_i^2 v_i(0)^2 \right\rangle }.
\label{eq:PF_discrete}
\end{equation}}

\gj{The first order Backward scheme can now be derived by approximating the integral using the rectangle method, leading to
\begin{equation}
\tilde{F}_i^{\overline{t+\Delta t}} = \tilde{F}_i^{\overline{t}} + F_i(t) \alpha(t) \Delta t +\mathcal{O}(\Delta t^2),
\end{equation}
with 
\begin{equation}
\alpha(t) = \frac{\left\langle M_i v_i(t) \tilde{F}_i^{\overline{t}} \right\rangle }{\left\langle M_i^2 v_i(0)^2 \right\rangle },
\end{equation}
and $ \tilde{F}_i^{\overline{0}} = F_i(0) $.
The algorithm to evaluate this equation is described in detail in Ref.~\cite{Carof2014} Eqs.~(21)-(27).}

\gj{However, it is also possible to approximate Eq.~\ref{eq:PF_discrete} with the trapezoidal rule, similar to the derivation of the second order Forward scheme \cite{Lesnicki2016}:
\begin{eqnarray}
&\tilde{F}_i^{\overline{t+\Delta t}} &= \tilde{F}_i^{\overline{t}} + F_i(t) \frac{\left\langle M_i v_i(t) \tilde{F}_i^{\overline{t}} \right\rangle }{\left\langle M_i^2 v_i(0)^2 \right\rangle } \frac{\Delta t}{2} \label{PF_2nd}\\
+ &F_i(t+\Delta t)& \frac{\left\langle M_i v_i(t+\Delta t) \tilde{F}_i^{\overline{t+\Delta t}} \right\rangle }{\left\langle M_i^2 v_i(0)^2 \right\rangle } \frac{\Delta t}{2} +  \mathcal{O}(\Delta t^3). \nonumber
\end{eqnarray}
By taking the correlation on both sides of Eq.~\ref{PF_2nd} with $ M_i v_i(t+\Delta t) $ and \gj{inserting} the solution back into Eq.~\ref{PF_2nd}, we can finally write down the propagator for the second order Backward orthogonal dynamics:
\begin{eqnarray}
\tilde{F}_i^{\overline{t+\Delta t}} &=& \tilde{F}_i^{\overline{t}} + F_i(t) \alpha(t) \frac{\Delta t}{2}\\
 &+&  \frac{ F_i(t+\Delta t)}{1-\frac{\Delta t}{2} \kappa } \left[ \zeta(t) + \epsilon \alpha(t) \frac{\Delta t}{2} \right] \frac{\Delta t}{2}  + \mathcal{O}(\Delta t^3), \nonumber
\end{eqnarray}
with
\begin{eqnarray}
\zeta(t) &=& \frac{\left\langle M_i v_i(t+\Delta t) \tilde{F}_i^{\overline{t}} \right\rangle }{\left\langle M_i^2 v_i(0)^2 \right\rangle },\\
\kappa &=& \frac{\left\langle M_i v_i(t) F_i(t) \right\rangle }{\left\langle M_i^2 v_i(0)^2 \right\rangle },\\
\epsilon &=& \frac{\left\langle M_i v_i(t+\Delta t) F_i(t) \right\rangle }{\left\langle M_i^2 v_i(0)^2 \right\rangle }.
\end{eqnarray}
\begin{figure}[b]
\includegraphics{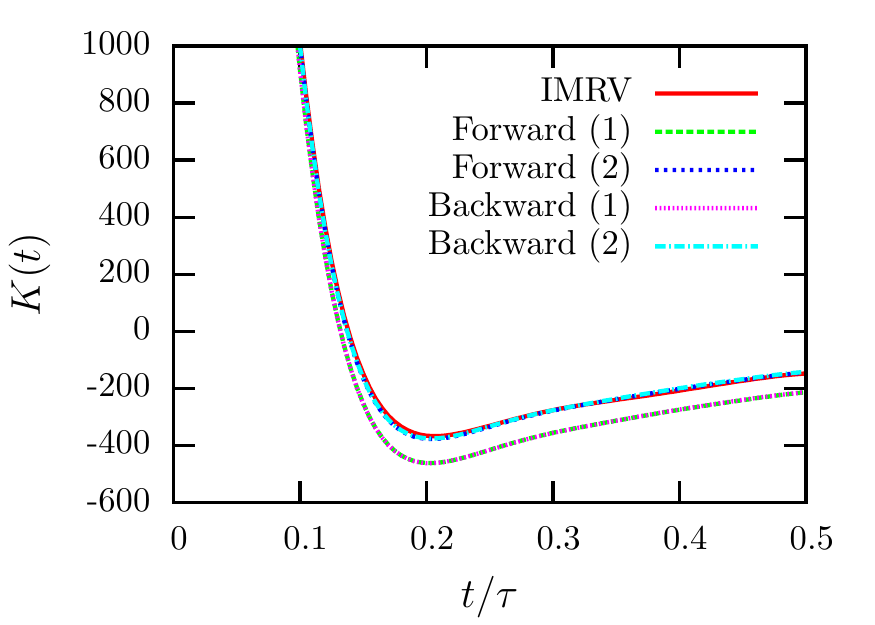}
\caption{\gj{Comparison of memory kernels reconstructed with orthogonal dynamics algorithms of first (1) and second (2) order. The time discretization was $ \Delta t_\text{MD} = 0.005 $. The IMRV result is included from Fig.~\ref{fig:iterative_comparison} as reference. The Forward (1) and Backward (1) as well as Forward (2) and Backward (2) curves lie exactly on top of each other.  }}
\label{fig:comparison_orthogonal}
\end{figure}
A numerical comparison between the presented extension to the Backward orthogonal dynamics and the already established Forward orthogonal dynamics can be found in Fig.~\ref{fig:comparison_orthogonal}. As expected, the second order algorithm performs much better than the first order scheme. Additionally it can be shown that the results of the two orthogonal algorithms are numerically identical. The Backward scheme, however, has the advantage to be applicable on-the-fly.}

\bibliography{library,library-control}

\providecommand{\latin}[1]{#1}
\providecommand*\mcitethebibliography{\thebibliography}
\csname @ifundefined\endcsname{endmcitethebibliography}
  {\let\endmcitethebibliography\endthebibliography}{}
\begin{mcitethebibliography}{32}
\providecommand*\natexlab[1]{#1}
\providecommand*\mciteSetBstSublistMode[1]{}
\providecommand*\mciteSetBstMaxWidthForm[2]{}
\providecommand*\mciteBstWouldAddEndPuncttrue
  {\def\EndOfBibitem{\unskip.}}
\providecommand*\mciteBstWouldAddEndPunctfalse
  {\let\EndOfBibitem\relax}
\providecommand*\mciteSetBstMidEndSepPunct[3]{}
\providecommand*\mciteSetBstSublistLabelBeginEnd[3]{}
\providecommand*\EndOfBibitem{}
\mciteSetBstSublistMode{f}
\mciteSetBstMaxWidthForm{subitem}{(\alph{mcitesubitemcount})}
\mciteSetBstSublistLabelBeginEnd
  {\mcitemaxwidthsubitemform\space}
  {\relax}
  {\relax}

\bibitem[Hij{\'{o}}n \latin{et~al.}(2010)Hij{\'{o}}n, Espa{\~{n}}ol,
  Vanden-Eijnden, and Delgado-Buscalioni]{Hijon2010}
Hij{\'{o}}n,~C.; Espa{\~{n}}ol,~P.; Vanden-Eijnden,~E.; Delgado-Buscalioni,~R.
  {Mori-Zwanzig formalism as a practical computational tool}. \emph{Faraday
  Discuss.} \textbf{2010}, \emph{144}, 301--322\relax
\mciteBstWouldAddEndPuncttrue
\mciteSetBstMidEndSepPunct{\mcitedefaultmidpunct}
{\mcitedefaultendpunct}{\mcitedefaultseppunct}\relax
\EndOfBibitem
\bibitem[Li \latin{et~al.}(2015)Li, Bian, Li, and Karniadakis]{Li2015}
Li,~Z.; Bian,~X.; Li,~X.; Karniadakis,~G.~E. {Incorporation of memory effects
  in coarse-grained modeling via the Mori-Zwanzig formalism}. \emph{J. Chem.
  Phys.} \textbf{2015}, \emph{143}, 243128\relax
\mciteBstWouldAddEndPuncttrue
\mciteSetBstMidEndSepPunct{\mcitedefaultmidpunct}
{\mcitedefaultendpunct}{\mcitedefaultseppunct}\relax
\EndOfBibitem
\bibitem[Li \latin{et~al.}(2017)Li, Lee, Darve, and Karniadakis]{Li2017}
Li,~Z.; Lee,~H.~S.; Darve,~E.; Karniadakis,~G.~E. {Computing the non-Markovian
  coarse-grained interactions derived from the Mori-Zwanzig formalism in
  molecular systems: Application to polymer melts}. \emph{J. Chem. Phys.}
  \textbf{2017}, \emph{146}, 014104\relax
\mciteBstWouldAddEndPuncttrue
\mciteSetBstMidEndSepPunct{\mcitedefaultmidpunct}
{\mcitedefaultendpunct}{\mcitedefaultseppunct}\relax
\EndOfBibitem
\bibitem[Shin \latin{et~al.}(2010)Shin, Kim, Talkner, and Lee]{Shin2010}
Shin,~H.~K.; Kim,~C.; Talkner,~P.; Lee,~E.~K. {Brownian motion from molecular
  dynamics}. \emph{Chem. Phys.} \textbf{2010}, \emph{375}, 316--326\relax
\mciteBstWouldAddEndPuncttrue
\mciteSetBstMidEndSepPunct{\mcitedefaultmidpunct}
{\mcitedefaultendpunct}{\mcitedefaultseppunct}\relax
\EndOfBibitem
\bibitem[Zwanzig(1961)]{Zwanzig1961}
Zwanzig,~R. {Memory Effects in Irreversible Thermodynamics}. \emph{Phys. Rev.}
  \textbf{1961}, \emph{124}, 983--992\relax
\mciteBstWouldAddEndPuncttrue
\mciteSetBstMidEndSepPunct{\mcitedefaultmidpunct}
{\mcitedefaultendpunct}{\mcitedefaultseppunct}\relax
\EndOfBibitem
\bibitem[Mori(1965)]{Mori1965}
Mori,~H. {Transport, Collective Motion, and Brownian Motion}. \emph{Prog.
  Theor. Phys.} \textbf{1965}, \emph{33}, 423--455\relax
\mciteBstWouldAddEndPuncttrue
\mciteSetBstMidEndSepPunct{\mcitedefaultmidpunct}
{\mcitedefaultendpunct}{\mcitedefaultseppunct}\relax
\EndOfBibitem
\bibitem[Zwanzig(2001)]{Zwanzig2001}
Zwanzig,~R. \emph{{Nonequilibrium statistical mechanics}}; Oxford University
  Press, 2001\relax
\mciteBstWouldAddEndPuncttrue
\mciteSetBstMidEndSepPunct{\mcitedefaultmidpunct}
{\mcitedefaultendpunct}{\mcitedefaultseppunct}\relax
\EndOfBibitem
\bibitem[Kinjo and Hyodo(2007)Kinjo, and Hyodo]{Kinjo2007}
Kinjo,~T.; Hyodo,~S.-a. {Equation of motion for coarse-grained simulation based
  on microscopic description}. \emph{Phys. Rev. E} \textbf{2007}, \emph{75},
  051109\relax
\mciteBstWouldAddEndPuncttrue
\mciteSetBstMidEndSepPunct{\mcitedefaultmidpunct}
{\mcitedefaultendpunct}{\mcitedefaultseppunct}\relax
\EndOfBibitem
\bibitem[Kubo(1966)]{Kubo1966}
Kubo,~R. {The fluctuation-dissipation theorem}. \emph{Reports Prog. Phys.}
  \textbf{1966}, \emph{29}, 306\relax
\mciteBstWouldAddEndPuncttrue
\mciteSetBstMidEndSepPunct{\mcitedefaultmidpunct}
{\mcitedefaultendpunct}{\mcitedefaultseppunct}\relax
\EndOfBibitem
\bibitem[Ceriotti \latin{et~al.}(2010)Ceriotti, Bussi, and
  Parrinello]{Ceriotti2010}
Ceriotti,~M.; Bussi,~G.; Parrinello,~M. {Colored-Noise Thermostats {\`{a}} la
  Carte}. \emph{J. Chem. Theory Comput.} \textbf{2010}, \emph{6},
  1170--1180\relax
\mciteBstWouldAddEndPuncttrue
\mciteSetBstMidEndSepPunct{\mcitedefaultmidpunct}
{\mcitedefaultendpunct}{\mcitedefaultseppunct}\relax
\EndOfBibitem
\bibitem[Carof \latin{et~al.}(2014)Carof, Marry, Salanne, Hansen, Turq, and
  Rotenberg]{Carof2014a}
Carof,~A.; Marry,~V.; Salanne,~M.; Hansen,~J.-P.; Turq,~P.; Rotenberg,~B.
  {Coarse graining the dynamics of nano-confined solutes: the case of ions in
  clays}. \emph{Mol. Simul.} \textbf{2014}, \emph{40}, 237--244\relax
\mciteBstWouldAddEndPuncttrue
\mciteSetBstMidEndSepPunct{\mcitedefaultmidpunct}
{\mcitedefaultendpunct}{\mcitedefaultseppunct}\relax
\EndOfBibitem
\bibitem[C{\'{o}}rdoba \latin{et~al.}(2012)C{\'{o}}rdoba, Indei, and
  Schieber]{Cordoba2012}
C{\'{o}}rdoba,~A.; Indei,~T.; Schieber,~J.~D. {Elimination of inertia from a
  Generalized Langevin Equation: Applications to microbead rheology modeling
  and data analysis}. \emph{J. Rheol.} \textbf{2012}, \emph{56}, 185--212\relax
\mciteBstWouldAddEndPuncttrue
\mciteSetBstMidEndSepPunct{\mcitedefaultmidpunct}
{\mcitedefaultendpunct}{\mcitedefaultseppunct}\relax
\EndOfBibitem
\bibitem[Guenza(1999)]{Guenza1999}
Guenza,~M. {Many chain correlated dynamics in polymer fluids}. \emph{J. Chem.
  Phys.} \textbf{1999}, \emph{110}, 7574--7588\relax
\mciteBstWouldAddEndPuncttrue
\mciteSetBstMidEndSepPunct{\mcitedefaultmidpunct}
{\mcitedefaultendpunct}{\mcitedefaultseppunct}\relax
\EndOfBibitem
\bibitem[C{\'{o}}rdoba \latin{et~al.}(2012)C{\'{o}}rdoba, Schieber, and
  Indei]{Cordoba2012a}
C{\'{o}}rdoba,~A.; Schieber,~J.~D.; Indei,~T. {The effects of hydrodynamic
  interaction and inertia in determining the high-frequency dynamic modulus of
  a viscoelastic fluid with two-point passive microrheology}. \emph{Phys.
  Fluids} \textbf{2012}, \emph{24}, 073103\relax
\mciteBstWouldAddEndPuncttrue
\mciteSetBstMidEndSepPunct{\mcitedefaultmidpunct}
{\mcitedefaultendpunct}{\mcitedefaultseppunct}\relax
\EndOfBibitem
\bibitem[Carof \latin{et~al.}(2014)Carof, Vuilleumier, and
  Rotenberg]{Carof2014}
Carof,~A.; Vuilleumier,~R.; Rotenberg,~B. {Two algorithms to compute projected
  correlation functions in molecular dynamics simulations}. \emph{J. Chem.
  Phys.} \textbf{2014}, \emph{140}, 124103\relax
\mciteBstWouldAddEndPuncttrue
\mciteSetBstMidEndSepPunct{\mcitedefaultmidpunct}
{\mcitedefaultendpunct}{\mcitedefaultseppunct}\relax
\EndOfBibitem
\bibitem[Lesnicki \latin{et~al.}(2016)Lesnicki, Vuilleumier, Carof, and
  Rotenberg]{Lesnicki2016}
Lesnicki,~D.; Vuilleumier,~R.; Carof,~A.; Rotenberg,~B. {Molecular
  Hydrodynamics from Memory Kernels}. \emph{Phys. Rev. Lett.} \textbf{2016},
  \emph{116}, 147804\relax
\mciteBstWouldAddEndPuncttrue
\mciteSetBstMidEndSepPunct{\mcitedefaultmidpunct}
{\mcitedefaultendpunct}{\mcitedefaultseppunct}\relax
\EndOfBibitem
\bibitem[Schnurr \latin{et~al.}(1997)Schnurr, Gittes, MacKintosh, and
  Schmidt]{B.Schnurr1997}
Schnurr,~B.; Gittes,~F.; MacKintosh,~F.~C.; Schmidt,~C.~F. {Determining
  Microscopic Viscoelasticity in Flexible and Semiflexible Polymer Networks
  from Thermal Fluctuations}. \emph{Macromolecules} \textbf{1997}, \emph{30},
  7781--7792\relax
\mciteBstWouldAddEndPuncttrue
\mciteSetBstMidEndSepPunct{\mcitedefaultmidpunct}
{\mcitedefaultendpunct}{\mcitedefaultseppunct}\relax
\EndOfBibitem
\bibitem[Lei \latin{et~al.}(2016)Lei, Baker, and Li]{Lei2016}
Lei,~H.; Baker,~N.~A.; Li,~X. {Data-driven parameterization of the generalized
  Langevin equation}. \emph{Proc. Natl. Acad. Sci.} \textbf{2016}, \emph{113},
  14183--14188\relax
\mciteBstWouldAddEndPuncttrue
\mciteSetBstMidEndSepPunct{\mcitedefaultmidpunct}
{\mcitedefaultendpunct}{\mcitedefaultseppunct}\relax
\EndOfBibitem
\bibitem[Fricks \latin{et~al.}(2009)Fricks, Yao, Elston, and
  Forest]{Fricks2009}
Fricks,~J.; Yao,~L.; Elston,~T.~C.; Forest,~M.~G. {Time-Domain Methods for
  Diffusive Transport in Soft Matter}. \emph{SIAM J. Appl. Math.}
  \textbf{2009}, \emph{69}, 1277--1308\relax
\mciteBstWouldAddEndPuncttrue
\mciteSetBstMidEndSepPunct{\mcitedefaultmidpunct}
{\mcitedefaultendpunct}{\mcitedefaultseppunct}\relax
\EndOfBibitem
\bibitem[Lyubartsev and Laaksonen(1995)Lyubartsev, and
  Laaksonen]{Lyubartsev1995}
Lyubartsev,~A.~P.; Laaksonen,~A. {Calculation of effective interaction
  potentials from radial distribution functions: A reverse Monte Carlo
  approach}. \emph{Phys. Rev. E} \textbf{1995}, \emph{52}, 3730--3737\relax
\mciteBstWouldAddEndPuncttrue
\mciteSetBstMidEndSepPunct{\mcitedefaultmidpunct}
{\mcitedefaultendpunct}{\mcitedefaultseppunct}\relax
\EndOfBibitem
\bibitem[Reith \latin{et~al.}(2003)Reith, P{\"{u}}tz, and
  M{\"{u}}ller-Plathe]{Reith2003}
Reith,~D.; P{\"{u}}tz,~M.; M{\"{u}}ller-Plathe,~F. {Deriving effective
  mesoscale potentials from atomistic simulations}. \emph{J. Comput. Chem.}
  \textbf{2003}, \emph{24}, 1624--1636\relax
\mciteBstWouldAddEndPuncttrue
\mciteSetBstMidEndSepPunct{\mcitedefaultmidpunct}
{\mcitedefaultendpunct}{\mcitedefaultseppunct}\relax
\EndOfBibitem
\bibitem[Gr{\o}nbech-Jensen and Farago(2013)Gr{\o}nbech-Jensen, and
  Farago]{Grønbech-Jensen2012}
Gr{\o}nbech-Jensen,~N.; Farago,~O. {A simple and effective Verlet-type
  algorithm for simulating Langevin dynamics}. \emph{Mol. Phys.} \textbf{2013},
  \emph{111}, 983--991\relax
\mciteBstWouldAddEndPuncttrue
\mciteSetBstMidEndSepPunct{\mcitedefaultmidpunct}
{\mcitedefaultendpunct}{\mcitedefaultseppunct}\relax
\EndOfBibitem
\bibitem[Br{\"{u}}nger \latin{et~al.}(1984)Br{\"{u}}nger, Brooks, and
  Karplus]{Brunger1984}
Br{\"{u}}nger,~A.; Brooks,~C.~L.; Karplus,~M. {Stochastic boundary conditions
  for molecular dynamics simulations of ST2 water}. \emph{Chem. Phys. Lett.}
  \textbf{1984}, \emph{105}, 495--500\relax
\mciteBstWouldAddEndPuncttrue
\mciteSetBstMidEndSepPunct{\mcitedefaultmidpunct}
{\mcitedefaultendpunct}{\mcitedefaultseppunct}\relax
\EndOfBibitem
\bibitem[Lee and Kapral(2005)Lee, and Kapral]{Lee2005}
Lee,~S.~H.; Kapral,~R. {Two-particle friction in a mesoscopic solvent}.
  \emph{J. Chem. Phys.} \textbf{2005}, \emph{122}, 214916\relax
\mciteBstWouldAddEndPuncttrue
\mciteSetBstMidEndSepPunct{\mcitedefaultmidpunct}
{\mcitedefaultendpunct}{\mcitedefaultseppunct}\relax
\EndOfBibitem
\bibitem[Theers \latin{et~al.}(2016)Theers, Westphal, Gompper, and
  Winkler]{Theers2016}
Theers,~M.; Westphal,~E.; Gompper,~G.; Winkler,~R.~G. {From local to
  hydrodynamic friction in Brownian motion: A multiparticle collision dynamics
  simulation study}. \emph{Phys. Rev. E} \textbf{2016}, \emph{93}, 032604\relax
\mciteBstWouldAddEndPuncttrue
\mciteSetBstMidEndSepPunct{\mcitedefaultmidpunct}
{\mcitedefaultendpunct}{\mcitedefaultseppunct}\relax
\EndOfBibitem
\bibitem[Hauge and Martin-L{\"{o}}f(1973)Hauge, and
  Martin-L{\"{o}}f]{Hauge1973}
Hauge,~E.~H.; Martin-L{\"{o}}f,~A. {Fluctuating hydrodynamics and Brownian
  motion}. \emph{J. Stat. Phys.} \textbf{1973}, \emph{7}, 259--281\relax
\mciteBstWouldAddEndPuncttrue
\mciteSetBstMidEndSepPunct{\mcitedefaultmidpunct}
{\mcitedefaultendpunct}{\mcitedefaultseppunct}\relax
\EndOfBibitem
\bibitem[Thalmann and Farago(2007)Thalmann, and Farago]{Thalmann2007}
Thalmann,~F.; Farago,~J. {Trotter derivation of algorithms for Brownian and
  dissipative particle dynamics}. \emph{J. Chem. Phys.} \textbf{2007},
  \emph{127}, 124109\relax
\mciteBstWouldAddEndPuncttrue
\mciteSetBstMidEndSepPunct{\mcitedefaultmidpunct}
{\mcitedefaultendpunct}{\mcitedefaultseppunct}\relax
\EndOfBibitem
\bibitem[Ricci and Cicotti(2003)Ricci, and Cicotti]{Ricci2003}
Ricci,~A.; Cicotti,~G. {Algorithms for Brownian dynamics}. \emph{Mol. Phys.}
  \textbf{2003}, \emph{101}, 1927--1931\relax
\mciteBstWouldAddEndPuncttrue
\mciteSetBstMidEndSepPunct{\mcitedefaultmidpunct}
{\mcitedefaultendpunct}{\mcitedefaultseppunct}\relax
\EndOfBibitem
\bibitem[Barrat and Rodney(2011)Barrat, and Rodney]{Barrat2011}
Barrat,~J.-L.; Rodney,~D. {Portable Implementation of a Quantum Thermal Bath
  for Molecular Dynamics Simulations}. \emph{J. Stat. Phys.} \textbf{2011},
  \emph{1}, 679--689\relax
\mciteBstWouldAddEndPuncttrue
\mciteSetBstMidEndSepPunct{\mcitedefaultmidpunct}
{\mcitedefaultendpunct}{\mcitedefaultseppunct}\relax
\EndOfBibitem
\bibitem[Plimpton(1995)]{Plimpton1995}
Plimpton,~S. {Fast Parallel Algorithms for Short-Range Molecular Dynamics}.
  \emph{J. Comput. Phys.} \textbf{1995}, \emph{117}, 1--19\relax
\mciteBstWouldAddEndPuncttrue
\mciteSetBstMidEndSepPunct{\mcitedefaultmidpunct}
{\mcitedefaultendpunct}{\mcitedefaultseppunct}\relax
\EndOfBibitem
\bibitem[Villa \latin{et~al.}(2010)Villa, Peter, and van~der Vegt]{Villa}
Villa,~A.; Peter,~C.; van~der Vegt,~N. F.~A. {Transferability of Nonbonded
  Interaction Potentials for Coarse-Grained Simulations: Benzene in Water}.
  \emph{J. Chem. Theory Comput.} \textbf{2010}, \emph{6}, 2434--2444\relax
\mciteBstWouldAddEndPuncttrue
\mciteSetBstMidEndSepPunct{\mcitedefaultmidpunct}
{\mcitedefaultendpunct}{\mcitedefaultseppunct}\relax
\EndOfBibitem
\end{mcitethebibliography}

\end{document}